\documentstyle[12pt]{article}

\newcommand{\be}{\begin{equation}}
\newcommand{\ee}{\end{equation}}

\newcommand{\bea}{\begin{eqnarray}}
\newcommand{\eea}{\end{eqnarray}}

\newcommand{\p}{\partial}

\newcommand{\nn}{\nonumber \\}
\newcommand{\f}{\frac}
\newcommand{\w}{\wedge}

\begin{document}
\thispagestyle{empty}

\begin{flushright}
{\bf arXiv: 0901.0599}
\end{flushright}
\begin{center} \noindent \Large \bf
Anisotropic gravity solutions in AdS/CMT
\end{center}

\bigskip\bigskip\bigskip
\vskip 0.5cm
\begin{center}
{ \normalsize \bf  Shesansu Sekhar Pal}\footnote{
Permanent address: {\sf Barchana, Jajpur, 754081, Orissa, India}}
\vskip 0.5cm

\vskip 0.5 cm
Saha Institute of Nuclear Physics, \\
1/AF, Bidhannagar, Kolkata 700 064, India\\
\vskip 0.5 cm
\sf shesansu${\frame{\shortstack{AT}}}$gmail.com
\end{center}
\centerline{\bf \small Abstract}

We have constructed gravity solutions by  breaking  the Lorentzian symmetry to 
its subgroup,  
which means there is Galilean symmetry but without the  rotational and 
boost invariance. 
This solution   shows anisotropic behavior along both the temporal and 
spatial directions as well as among the spatial directions 
and more interestingly, it displays the precise 
scaling symmetry 
required for metric as well as the form fields. From the field theory point
of view, it describes a  theory 
which respects the scaling  symmetry, 
$t\rightarrow \lambda^{z_1}t,~x\rightarrow \lambda^{z_2}t,~
y\rightarrow \lambda y$, for $z_1\neq z_2$, as well as the translational 
symmetry associated to both time and space directions, which means we have 
found a non-rotational but
Lifshitz-like fixed points from the dual field theory point of view. We also 
discuss the minimum number of generators required to see the appearance of 
such Lifshitz points. In $1+1$ dimensional field theory, 
it is $3$ and for $2+1$ dimensional field theory, the number is $4$.

\newpage
\section{Introduction and motivation}

After the recent of advancement of gauge-gravity duality conjecture 
\cite{jmm}, \cite{gkp} and \cite{ew}, 
it has been becoming very interesting to apply this duality conjecture to 
somewhat realistic situations that is in condensed matter theories, so as to 
understand more about it, in the sense to have a better understanding of 
quantum  phase diagram \cite{ss} and \cite{sm}, 
at the strong coupling limit. In this context, it has 
been proposed in \cite{klm} that there exists a gravity solution for the 
Lifshitz-like fixed point and latter it was generalized to gravity solution 
showing only the temporal scale invariance in \cite{ssp} and to solutions 
in arbitrary space time dimensions using the p-form fields in \cite{ssp1}. 
In particular, 
the solution in $d$ space time dimension is constructed so as to have 
a global $O(d-2)$ symmetry
 and in \cite{mt} a proper holographic renormalization prescription is 
initiated to compute the  correlation functions and some other studies 
in \cite{ph}.

It is interesting to recall that there are many important studies has been 
done in the context of quantum phase transitions using global 
rotational symmetry in field space.
For example the super fluid-insulator transition in $\phi^4$ theory with $O(2)$ 
symmetry in 2+1 space time theory, but with a Lorentzian symmetry, more 
appropriately with a relativistic conformal invariance  \cite{gmehb} and 
\cite{spp}, in arbitrary dimensions in \cite{fwgf} and some other studies 
related to quantum critical points in 
\cite{hls}, \cite{gg}, \cite{vbs} and \cite{svbsf}.

The signature of quantum phase transition can be attributed to, from 
\cite{ss}, as  the appearance of non-analyticity in the ground state energy
as a function of a dimensionless parameter $g$ at $g=g_c$ and at 
zero temperature. However, due to the unavoidable thermal fluctuations  the
system is studied around $|g-g_c|$ and at finite temperature, $T.$  The 
mathematical definition for it is, by finding the quantum fluctuations to 
energy, $\Delta$, at zero T as a function of the parameter $g$
\be
\Delta\sim J |g-g_c|^{z\nu},
\ee 
where $J$ is the energy scale of a characteristic microscopic coupling and
$z\nu$ is the universal  critical exponent. The other important ingredient is 
the characteristic correlation length $\xi$, which is related to $\Delta$ as
\be
\Delta~\sim~\xi^{-z}.
\ee
It is important to note that both $\nu$ and $z$ are positive as 
 $\xi$ diverges at the critical point and  $\Delta$ vanishes at the 
critical point $g_c.$

The Lifshitz point is described as a multicritical point on the curve 
describing the $\lambda$ line of the second order phase transition \cite{hls},
which  separates ordered phases. For example, in a magnetic systems the 
Lifshitz point is a triple point where paramagnetic, ferromagnetic and 
helicoidal phases meet \cite{bsoc}.

Let us recall few definitions from \cite{mh},  for a scaling transformation 
under which the correlation function $C(x_i,t)$ obeys
\be
C(\lambda x_i;~\lambda^z~t)=\lambda^{-2 \delta} C(x_i;t),
\ee
where $\delta$ is the scaling dimension and $z$ is called the anisotropy 
exponent or the dynamical exponent in equilibrium phase transition, 
especially for systems having $z\neq 1$ is called as strongly anisotropy 
critical systems. Examples of where the strongly anisotropy arises are 
(a) systems exhibiting percolation \cite{wk} and (b) systems showing the 
appearance of Lifshitz points \cite{hls}.

For $z=1$, we have already witnessed the examples that has the features of  
being recognized as the quantum critical points, these are the $AdS_d$ 
space times which respects the full conformal algebra. Whereas for $z=2$, there
were systems constructed very recently, which has the reduced symmetry group
and is called as the Schr$\ddot{o}$dinger group. It is interesting to note 
that the 
critical points of $z=1$ theories are Lorentz invariants whereas for higher
values of $z$, the theory should necessarily break the Lorentz symmetry.

There is also a  lot of other activity in a somewhat related area starting 
from \cite{son} and  \cite{bm}, where the symmetry is enhanced than  
that we study here and is called as Schr$\ddot{o}$dinger group. 
It contains the generators for 
spatial translations, time translations, rotations, boosts, number operator
and dilatation operators. Latter this solution was embedded in string theory 
in \cite{hrr},\cite{mmt} and \cite{abm} using either the Null Melvin Twist 
\cite{ghhlr}, \cite{ag} or the TsT symmetry prescriptions \cite{ml} and 
this is applied further in \cite{ssp2} for the Sakai-Sugimoto model and some 
other examples in \cite{mot}, also in \cite{aadv} and there are some other
related but earlier studies in \cite{ch},\cite{msw} and some recent studies in
\cite{ns}-\cite{mp}. Also some aspects of non-relativistic hydrodynamics
is studied in \cite{rrst}.

In this paper we have generated  some other interesting solutions in the 
bulk by taking gravity and
form-fields as the relevant degrees of freedom with Chern-Simon interaction. 
In particular, in 2+1 
dimension by taking   gravity and a 2-form field strength, we have constructed 
solution that exhibits the presence of a dynamical exponent $z$ and this
solution has the continuous symmetries of time translation, space translation 
and scaling symmetry. The discrete symmetries are time and space reflections. 
This solution is generated using a combination of both electric and magnetic
2-from flux. In this solution, we can have only temporal scaling symmetry 
as well as both temporal and spatial scaling symmetry.

In 3+1 dimension, we generate solutions that possess time translation, spatial
translation and scaling symmetry but without any rotational or Galilean 
transformation, the analogue of boost symmetry. The discrete symmetries 
are the same as in 2+1 dimension.  More interestingly, this 
solution has got two non-trivial exponents, $z_1$ and $z_2$. This solution 
is constructed using gravity, a 2-from field strength and two 3-form 
fieldstrenghs, which includes a massive 2-form potential. 
     
The solutions that we have constructed in this paper uses an effective action
in the bulk which tells us that we can have quantum critical points starting 
from  $1+1$ field theory and to see this we do not need to have the full 
symmetry group i.e. either the Lorentzian conformal symmetry or the 
Schr$\ddot{o}$dinger group, with a lots of generators, instead a minimum of 
$3$ generators for this particular space time, time translation, spatial translation and scaling symmetry, can generate the required dual solution.
  
The plan of the paper is in section 2, we shall study the $2+1$ dimensional 
bulk theory and in section 3, the $3+1$ dimensional theory and finally 
conclude in section 4.

\section{1+1 dim field theory systems}
In 2+1 dimensional bulk theory the solution, recalling from \cite{ssp1},
\be\label{config1}
ds^2=L^2[-r^{2a}dt^2+dx^2+\f{dr^2}{r^2}],~~~F_2=\f{AL^2}{r^{1-a}}dr\w dt,~~~
A^2=\f{2a^2}{L^2},~~~\Lambda=-\f{a^2}{2L^2},
\ee 

which only shows the temporal scale invariance and is one of the  allowed 
solution at zero temperature containing  gravity and $F_2$-from  
 as the non-trivial degrees of freedom.     
The  other possible solution, more appropriately with these degrees of freedom,  in 2+1 space time dimension  can be  realized  by taking  the action 
\be
S=\f{1}{2\kappa^2}\int dx dr dt \bigg[\sqrt{-g}\bigg(R-2\Lambda-
\f{F^2_2}{4}-\f{H^2_2}{4}\bigg)-c~\epsilon^{MNP}{\cal A}_M\p_N {\cal B}_P\bigg]
\ee
with the ansatz to fluxes as 
\be
F_2=d{\cal A}=A_1 L^2 r^{a-1}dr\w dt,~~~H_2=d{\cal B}=A_2 L^2 r^{b-1} dr\w dx
\ee
gives the solution as

\bea\label{config2}
ds^2&=&L^2[-r^{2a}dt^2+r^{2b}dx^2+\f{dr^2}{r^2}],~~~A^2_2=\f{2b(a-b)}{L^2},~~~
A^2_1=\f{2a(a-b)}{L^2},\nn 
c^2L^2&=&ab,~~~\Lambda=-\f{a^2+b^2}{2L^2}
\eea
where $b\neq 0$ and $a~\geq~b~>~0.$ 

By doing a change of coordinates, and defining $z:=\f{a}{b}$, the metric becomes\be\label{2+1_background}
ds^2=L^2[-\rho^{2z}dt^2+\rho^2dx^2+\f{d\rho^2}{\rho^2}]
\ee


The generators that follows from eq(\ref{2+1_background}) are
\be
H=-i\p_t,~~~P_x=-i\p_x,~~~D=-i[-zt\p_t-x\p_x+\rho\p_{\rho}]
\ee
and the algebra it satisfy 
\be
[D,P_x]=-i~P_x,~~~[D,H]=-iz~H
\ee
\subsection{Dimensions and Correlators}
In this subsection, we would like to consider a massive scalar field of mass,
 $m$  that is moving in the Euclidean background of 
eq(\ref{2+1_background}) and from
which we shall generalize the gauge gravity duality \cite{kw} and present the 
dictionary. In due
course we shall determine the dimensions of the operators that are dual to 
the bulk massive scalar field and present the conditions when and which
of the dimensions will become relevant.

The equations of motion of the scalar field, $\phi$, for a special value to 
the exponent, $z=2$  in $u=1/r$,  coordinate system 
\be
\p^2_u\phi-\f{2}{u}\p_u\phi-\bigg[u^2 w^2+{\bf k}^2+\f{m^2L^2}{u^2} \bigg]\phi=0
\ee

Asymptotically the operator $\Delta$, dual to massive scalar field $\phi$ 
satisfies
\be
\Delta(\Delta-3)=m^2 L^2,~~~\Delta_{\pm}=\f{3}{2}\pm\sqrt{\f{9}{4}+m^2L^2}.
\ee

The part of the generalized structure to gauge gravity correspondence follows 
from requiring the finiteness of the Euclidean action and is that if the 
mass of scalar field stays above the following bound 
\be
(m~L)^2~>~-\f{5}{4}
\ee
then only the $\Delta_+$ branch is allowed as the only possible solution. If 
the mass satisfies the following condition 
\be
-\f{9}{4}~<~(m~L)^2~<~-\f{5}{4}
\ee
then both the $\Delta_+$ and $\Delta_-$ branches are allowed. The analogue of 
 Breitenlohner-Freedman bound \cite{bf} is
\be 
(m~L)^2~>~-\f{9}{4},
\ee
which says if the mass of the scalar field stays below this bound then there is an 
instability in the system.

For computation of the two point correlation function involving operators that
are dual to the scalar field requires the form of the Green's function 
$G(u,k)$, which is related to the source $\phi(0,k)$ 
\be
\phi(u,k)=G(u,k)\phi(0,k).
\ee
Note that $k=(w,{\bf k})$ and 
the solution to the wave equation satisfied by the scalar field is
\bea
G(u,{\bf k})&=&c_1 2^{\f{2+\sqrt{9+4m^2L^2}}{4}}\times e^{-\f{wu^2}{2}}\times 
u^{\f{3+\sqrt{9+4m^2L^2}}{2}}\times,\nn && 
U\bigg(\f{{\bf k}^2+(2+\sqrt{9+4m^2L^2})w}{4w},\f{2+\sqrt{9+4m^2L^2}}{2}, wu^2\bigg)
\eea
where $U(a,b,z)$ is the confluent hypergeometric function of the second kind 
and $c_1$ is the normalization constant that need to be determined by 
imposing the following condition
\be
G(u\rightarrow\epsilon,{\bf k},w)=1
\ee

From, now onwards we shall be considering only the massless case and in this
case  the properly normalized propagator 
\be
G(u,{\bf k})=\f{2}{\sqrt{\pi}} \Gamma(\f{{\bf k}^2+5w}{4w}) e^{-\f{wu^2}{2}}
U\bigg(\f{{\bf k}^2-w}{4w},-\f{1}{2}, wu^2\bigg)
\ee

From the Euclidean action of the scalar field, upon using the equations of 
motion results
\be\label{action_ft1}
S=\int d{\bf k}dw \phi(0,-k) {\cal F}(k)\phi(0,k),
\ee
where the flux factor is
\be 
{\cal F}(k)=-2 \sqrt{g}g^{uu}\p_u G(u,k)|^{\infty}_{\epsilon}.
\ee

In order to calculate the flux factor, we need the expansion of the 
Green's function and its derivative up to  quartic order in $u$
\bea 
G(u,k)&=&1-\f{{\bf k}^2 u^2}{2}+\f{8}{3} w^{\f{3}{2}}
\f{\Gamma(\f{{\bf k}^2+5w}{4w})}{\Gamma(\f{{\bf k}^2-w}{4w})}u^3+\f{1}{8}(2w^2-{\bf k}^4)u^4+{\cal O}(u^5),\nn 
\p_uG(u,k)&=&-{\bf k}^2 u+8w^{\f{3}{2}}
\f{\Gamma(\f{5w+{\bf k}^2}{4w})}{\Gamma(\f{{\bf k}^2-w}{4w})}u^2+\f{2w^2-{\bf k}^4}{2}u^3+\f{4}{3} {\bf k}^2w^{\f{3}{2}}
\f{\Gamma(\f{5w+{\bf k}^2}{4w})}{\Gamma(\f{{\bf k}^2-w}{4w})}u^4+{\cal O}(u^5),\nn
\sqrt{g}g^{uu}=\f{L}{u^2}
\eea
 
Now including everything and dropping the divergent terms in the 
$\epsilon\rightarrow 0$ limit, gives us  the correlator
\be
<{\cal O}(-k){\cal O}(k)>={\cal F}(k)=-16 L  w^{\f{3}{2}}
\f{\Gamma(\f{{\bf k}^2+5w}{4w})}{\Gamma(\f{{\bf k}^2-w}{4w})}
\ee

This structure of the 2-point correlator matches exactly as is found in 
\cite{mt}, which means upon addition of proper counter terms to the action 
eq(\ref{action_ft1}) will help us to cancel the irrelevant divergent terms.


\section{3+1 dim: Symmetries}
Let us do some studies of  $3+1$ dimensional examples which has 
 got generators associated to time 
translations $(H)$, spatial translations $(P_x),~(P_y)$ and 
dilatation generator, $(D)$. The dilatation 
generator  associated to the scaling symmetry that we have has a very
 specific structure and  it's in the notation of \cite{ssp}
\be\label{scaling_symmetry}
t\rightarrow\lambda^a~t,~~~x\rightarrow\lambda^b~x,~~~
y\rightarrow\lambda^c~y,~~~r\rightarrow\f{r}{\lambda}.
\ee

The metric that shows the above symmetries can have a structure 
\be
ds^2=L^2[-r^{2a}dt^2+r^{2b}dx^2+r^{2c}dy^2+\f{dr^2}{r^2}],
\ee
where $L$ is the size of the space time and the exponents $a,~b,~c$ can take 
any real values. Let us do the following coordinate transformation for 
$c\neq 0$
\be
r^c =\rho,~~~(t,~x,~y)\rightarrow \f{1}{c}(t,~x,~y),~~~L\rightarrow L~c
\ee
and the resulting metric for $z_1:=\f{a}{c}$ and $z_2:=\f{b}{c}$
\be
ds^2=L^2[-\rho^{2z_1}dt^2+\rho^{2z_2}dx^2+\rho^2dy^2+\f{d\rho^2}{\rho^2}]
\ee

Using this form of metric, the explicit structure of the generators in 
generic case, $z_1\neq z_2$, are 
\be
H=-i \p_t,~~~P_x=-i\p_x,~~~P_y=-i\p_y,~~~D=-i[-z_1t\p_t-z_2x\p_x-y\p_y+\rho\p_{\rho}]
\ee
and these generators obey the following algebra
\be
[D,P_x]=-iz_2P_x,~~~[D,P_y]=-iP_y,~~~[D,H]=-iz_1H.
\ee

From the algebra or from the metric it follows that there exist two distinct 
nonequivalent exponents $z_1$ and $z_2$. For a special value, namely 
$z_1=z_2=1$, the algebra is enhanced to that of the $SO(2,3)$ of 
$AdS_4$ spacetime. It is interesting to note that these exponents can 
take any real values as long as they do not break any consistency of the 
theory e.g. the unitarity or the geodesically incompleteness of the space 
time.  

\subsection{Constraints}
It looks like that if the dual field theory  have an action like
\be\label{action_ft}
S=\f{1}{2}\int dt dx dy [-(\p_t\chi)^{\alpha}+K_1(\p^2_x\chi)^{\beta}+
K_2(\p^2_y\chi)^{\gamma}],
\ee
where $\chi$ is the field in the field theory with $K_1$ and $K_2$ are two 
constants and describes a plane containing 
fixed points. Roughly, speaking, there may be two constants from the 
dual field theory point of view as we have introduced  the  asymmetry
between the spatial directions $x$ and $y$.

Now if we demand that the above  action eq(\ref{action_ft}), 
respects the scaling 
symmetry eq(\ref{scaling_symmetry}), where the field $\chi$ transforms 
trivially, implies that the  $\alpha,~\beta,~\gamma$ should take the 
following values
\be
\alpha=1+\f{z_2}{z_1}+\f{1}{z_1},~~~\beta=\f{1}{2}+\f{z_1}{2z_2}+\f{1}{2z_2},~~~\gamma=\f{1+z_1+z_2}{2}.
\ee

Now, it is trivial to see that if we want the action eq(\ref{action_ft}) 
to be quadratic in fields, $\chi$ implies $z_1=2$ and 
$z_2=1$\footnote {In this context, it is probably correct to say that for a field theory which 
is a  unitary 
theory should have a dual bulk space time solution which obeys  the 
constraints of being a real solution with negative cosmological constant,  
 imposes the restrictions that we may not
 have a solution which shows only spatial scale invariance.}.

Let us study under what condition the field theory can be non-unitary ?
The way to see non-unitary behavior is to have the property of 2-point 
correlation function of  operators that  increases with their distance of 
separation
\be
<{\cal O}(x){\cal O}(0)>\sim \f{1}{|x|^{2\Delta}},
\ee 
where $\Delta$ is the dimension of the operator ${\cal O}(0),$ 
which is assumed to take only negative values.

For a minimally coupled scalar field, $\phi(x)$, of mass $m$ moving in the 
Euclidean background of 
\be 
ds^2=L^2[-r^{2a}dt^2+r^{2b}dx^2+r^{2c}dy^2+\f{dr^2}{r^2}]
\ee
obeys an equation
\be
\p^2_u\phi+\f{(1-a-b-c)}{u}\p_u\phi-\bigg[u^{2(a-1)}\omega^2+
u^{2(b-1)}k^2_x+u^{2(c-1)}k^2_y+\f{m^2L^2}{u^2} \bigg]\phi=0,
\ee
where $u=1/r.$
The structure of the generalized form of the field theory and gravity 
correspondence can be understood  by finding the relation between the mass
$m$ of the field, in this case a scalar field $\phi$, with the dimension of the
operator $\Delta,$  that it is associated to in the dual field theory.

In the case of interest i.e. for the massive scalar field the relation is
\bea\label{delta}
&&\Delta^2-(a+b+c)\Delta-m^2L^2=0,\nn
&&\Delta_{\pm}=\f{(a+b+c)}{2}\pm \sqrt{\bigg(\f{a+b+c}{2}\bigg)^2+m^2L^2},
\eea
where $\Delta_{\pm}$ are the two roots and $\Delta_+~>~\Delta_-$.

The analogue of Breitenlohner-Freedman bound \cite{bf} is
\be
(mL)^2 > -\bigg(\f{a+b+c}{2}\bigg)^2.
\ee

In order to see the situation, when the 2-point correlation function between 
the operators dual to scalar field increases, i.e. when the conformal 
dimension 
$\Delta$ becomes negative. The easiest way is to take the $\Delta_-$ branch 
and \be 
\Delta_-=\f{(a+b+c)}{2}- \sqrt{\bigg(\f{a+b+c}{2}\bigg)^2+m^2L^2}
\ee
from which it follows that if 
\be
a+b+c ~~~<~~~0,
\ee
then it is trivial to see that the dimension of operators is negative, as long 
as,  the object under the square-root is positive i.e. 
\be 
\bigg(\f{a+b+c}{2}\bigg)^2+(mL)^2~>0.
\ee
From which it follows that
 the 2-point correlation function increases with their distance of 
separation for the total negative sum of the exponents  and to recall 
that the $\Delta_-$ branch comes into picture 
when the mass of the scalar field obeys 
\be
-\bigg(\f{a+b+c}{2}\bigg)^2 ~<~ (mL)^2~< 1-\bigg(\f{a+b+c}{2}\bigg)^2
\ee
this condition. In fact, in this range of the $m^2$ both $\Delta_+$ and 
$\Delta_-$ branch comes into picture.

However, if the mass of the scalar field obeys
\be
m^2L^2~>1-\bigg(\f{a+b+c}{2}\bigg)^2 
\ee
then only $\Delta_+$ branch comes into picture and to see under what condition 
can $\Delta_+$ becomes negative, so as to have the non-unitary feature of the 
2-point correlation function ? 

Generically, it does not becomes negative, however if 
$\bigg(\f{a+b+c}{2}\bigg) $ dominates the term that comes under the 
square-root of $\Delta_+$, as it is always positive  then $\Delta_+$ can 
 becomes negative, i.e. if 
\be
a+b+c~<~0, ~~~ {\rm and}~~~ \bigg(\f{a+b+c}{2}\bigg)~>~\sqrt{\bigg(\f{a+b+c}{2}\bigg)^2+m^2L^2},
\ee
these conditions are full filled.

So, it just follows that if the sum of the exponents, $a+b+c$,
 is negative then 
the theory is not unitary and it  says that even if one or two of the exponents
are negative and if the total sum of the exponents is not negative then the 
theory is still unitary.

But this conclusion somehow does not make sense. Let us consider a situation 
where we take all the 
exponents as same and negative then we go over to $AdS_4$ spacetime, which is 
a unitary theory and contradicts with the conclusion of having a 
non-unitary theory for $a+b+c~<~0.$ 

So the proper way  to find the dimension of the
operators is to  go over to the $\rho$-coordinate system with the exponents
$z_1$ and $z_2$. However, for the computation of finding new solutions, the 
$r$-coordinate system is probably better and also to analyze  solutions with 
only temporal scale invariance.
 
The easiest way  to get the expression of the dimension of the massive
scalar field  is to start from eq(\ref{delta}) and define 
$ \f{\Delta}{c}$ as the new $\Delta$ and $\f{L}{c}$ as the new size parameter
$L$ and this is meaningful only when $c\neq 0.$ It gives
\bea
&&\Delta^2-(1+z_1+z_2)\Delta-m^2L^2=0,\nn
&&\Delta_{\pm}=\f{(1+z_1+z_2)}{2}\pm \sqrt{\bigg(\f{1+z_1+z_2}{2}\bigg)^2+m^2L^2},
\eea

Now the previous analysis of getting the negative dimension translates to 
\be
1+z_1+z_2~<~0.
\ee
So, we see that negativity  of one of the exponent do not necessarily  make 
the sum  $1+z_1+z_2$ to become negative, which means the correlation 
function of massive 
scalar fields  do not increases with their distance of separation even if one
of the exponent becomes negative.

However, if we make one of exponents negative, say $z_1$, then the time 
coordinate shrinks to zero at $\rho\rightarrow\infty$ i.e. 
at the boundary and this says that the space time is not geodesically 
complete at the boundary and the same is true for the other exponent $z_2.$

\subsection{The models in the bulk theory}
In order to discuss the complete anisotropy in a 3+1 dimensional theory
 between the temporal and spatial
coordinates as well as the anisotropy among  the spatial coordinates, we need
to consider an action which has the following degrees of freedom: metric, a 
two-form massive potential, a 2-form field strength and a 3-form 
field strength. In our discussion we have included the 4-form flux for 
completeness but it is not required and its absence is not going to change 
anything as roughly in 3+1 dimension, it is dual to a scalar field. 
Note that the
cosmological constant is already there in the effective gravitational action.

The form of  p-form fluxes need to be taken comes in a very specific way 
such that  
we need to have an anisotropy at the end of the calculation. Typically, the 
$B_2$ and $F_2$ form objects should be extended in a such a way that  they
have  one leg in common and two legs for the $H_3$ and $F_3$ form field 
strength objects. 

Let us look at a  model in 4 dimensional theory, with an  action 
\be 
\label{gen_action2}
 S= \f{1}{2\kappa^2}\int \bigg[\sqrt{-g}\bigg(R-2\Lambda-
\f{H^2_3}{12}-\f{m^2_0}{2} B^2_2-\f{F^2_3}{12}-\f{F^2_2}{4}-\f{F^2_4}{24}
\bigg)- c_1\epsilon^{M_1M_2M_3M_4} A_{M_1M_2}F_{M_3M_4}\bigg],
\ee
where $c_1$ is the topological coupling. 
The equations of motion that follows from it  are 
\bea
&&\p_{M_3}(\sqrt{-g} F^{M_3M_4})+\f{2}{3}\epsilon^{M_1M_2M_3M_4} 
c_1 F_{M_1M_2M_3}=0,\nn
&&\p_{M_3}(\sqrt{-g} H^{M_1M_2M_3})-2m^2_0 \sqrt{-g} B^{M_1M_2}=0,~~~
\p_{M_1}(\sqrt{-g} F^{M_1M_2M_3M_4})=0,\nn
&&\p_{M_3}(\sqrt{-g} F^{M_1M_2M_3})-2 c_1 \epsilon^{M_1M_2M_3M_4} F_{M_3M_4}=0,\nn  
&&R_{MN}-\f{1}{2}g_{MN}R+\f{1}{2}g_{MN}\bigg[2\Lambda+
\f{H^2_3}{12}+\f{m^2_0}{2} B^2_2+\f{F^2_3}{12}+\f{F^2_2}{4}+\f{F^2_4}{24}\bigg]-
\f{1}{4}H_{MM_1M_2}{H_N}^{M_1M_2}-\nn &&\f{1}{2}F_{MM_1}{F_N}^{M_1}
-m^2_0 B_{MM_1}{B_N}^{M_1}-\f{1}{4}F_{MM_1M_2}{F_N}^{M_1M_2}-\f{1}{6}F_{MM_1M_2M_3}{F_N}^{M_1M_2M_3}=0
\eea

Using an ansatz of the following  kind for the metric and form fields 

\bea\label{ansatz_2}
ds^2&=&L^2[-r^{2a}dt^2+r^{2b}dx^2+r^{2c}dy^2+\f{dr^2}{r^2}]\nn
B_2&=& A_2 L^2 r^{a+b} dt\w dx,~~~H_3=A_2(a+b)L^2 r^{a+b-1}dr\w dt\w dx,\nn
F_2&=&A_1 L^2 r^{a-1} dr\w dt,~~~F_3=B L^3 r^{b+c-1} dr\w dx\w dy,\nn
F_4&=& f_0 L^4 r^{a+b+c-1}dr\w dt\w dx\w dy
\eea

where $A_1,~A_2$ and $B$ are constants. Solving the equations of motion of the
$F_2$ and $F_3$ fluxes gives
\be
c_1 L=\f{(b+c)}{4B} A_1=\f{aB}{4A_1},
\ee
with the solutions to topological couplings $c_1$  as 
\be
16 c^2_1L^2=a(b+c),
\ee
 and the solution to $B_2$ equations of motion gives
\be\label{mass_b2}
2m^2_0L^2=c(a+b).
\ee

The equations of motion that results from the metric components are
\bea\label{metric_eom2}
&&4(b^2+bc+c^2)+(4\Lambda+2f^2_0+B^2+A^2_1 )L^2+A^2_2[(a+b)^2+2 m^2_0 L^2]=0,
\nn
&&4(a^2+ac+c^2)+(4\Lambda+2f^2_0-B^2-A^2_1 ) L^2+A^2_2[(a+b)^2+2 m^2_0 L^2]=0,
\nn
&&4(b^2+ba+a^2)+(4\Lambda+2f^2_0-B^2-A^2_1 ) L^2-A^2_2[(a+b)^2+2 m^2_0 L^2]=0,
\nn
&&4(ba+bc+ca)+(4\Lambda+2f^2_0-B^2+A^2_1 ) L^2+A^2_2[(a+b)^2-2 m^2_0 L^2]=0
\eea

Solving    the equations of eq(\ref{metric_eom2}), using eq(\ref{mass_b2}) 
gives
\bea\label{special_sol2}
A^2_1&=&\f{2a(a-b)}{L^2},~~~B^2=\f{2(a-b)(b+c)}{L^2},~~~
\Lambda=-\f{a^2+ab+2b^2+bc+c^2+f^2_0L^2}{2L^2},\nn
A^2_2&=&\f{2(b-c)}{(a+b)}
\eea

It just says that in order to have a real solution, we should have a constraint
as \be
a~~~\geq ~~~b~~~\geq~~~c~~~>~~~0,
\ee 

which means the exponents $a,~~~b$ and $c$ has to be positive.\\

Let us take  another choice to the ansatz for metric and form fields, 
instead of taking the ansatz as eq(\ref{ansatz_2}) and  if we  take

\bea\label{ansatz_3}
ds^2&=&L^2[-r^{2a}dt^2+r^{2b}dx^2+r^{2c}dy^2+\f{dr^2}{r^2}]\nn
B_2&=& A_2 L^2 r^{b+c} dx\w dy,~~~H_3=A_2(b+c)L^2 r^{b+c-1}dr\w dx\w dy,\nn
F_2&=&A_1 L^2 r^{b-1} dr\w dx,~~~F_3=B L^3 r^{a+c-1} dr\w dt\w dy,\nn
F_4&=& f_0 L^4 r^{a+b+c-1}dr\w dt\w dx\w dy
\eea

then the solutions to topological coupling, $c_1$ and $m^2_0$ as well as to 
$A_1,~A_2,~B$ and $\Lambda$ are
\bea
16 c^2_1 L^2&=&b(a+c),~~~2 m^2_0L^2=a(b+c),~~~A^2_1=\f{2b(c-b)}{L^2},~~~
A^2_2=\f{2(a-c)}{(b+c)},\nn
B^2&=& 2\f{(c-b)(a+c)}{L^2},~~~\Lambda=-\f{a^2+b^2+ac+bc+2c^2+f^2_0L^2}{2L^2}
\eea

This solution makes sense only when 
\be
a~~~\geq c~~~\geq~~~b~~~>~0,
\ee
which means all the exponents are positive and are real. Now we can do a simple
renaming of $b\leftrightarrow c$ and generate another solution. 
But, the consistent way to do so is to  take
 the following  ansatz for metric and form fields

\bea\label{ansatz_3}
ds^2&=&L^2[-r^{2a}dt^2+r^{2b}dx^2+r^{2c}dy^2+\f{dr^2}{r^2}]\nn
B_2&=& A_2 L^2 r^{b+c} dx\w dy,~~~H_3=A_2(b+c)L^2 r^{b+c-1}dr\w dx\w dy,\nn
F_2&=&A_1 L^2 r^{c-1} dr\w dy,~~~F_3=B L^3 r^{a+b-1} dr\w dt\w dx,\nn
F_4&=& f_0 L^4 r^{a+b+c-1}dr\w dt\w dx\w dy
\eea

then the solutions to topological coupling, $c_1$ and $m^2_0$ as well as to 
$A_1,~A_2,~B$ and $\Lambda$ are
\bea
16 c^2_1 L^2&=&c(a+b),~~~2 m^2_0L^2=a(b+c),~~~A^2_1=\f{2c(b-c)}{L^2},~~~
A^2_2=\f{2(a-b)}{(b+c)},\nn
B^2&=& 2\f{(b-c)(a+b)}{L^2},~~~\Lambda=-\f{a^2+c^2+ab+bc+2b^2+f^2_0L^2}{2L^2}
\eea 

This solution makes sense only when 
\be
a~~~\geq b~~~\geq~~~c~~~>~0.
\ee

So, we have generated  several solutions with the desired anisotropy among the 
coordinates.

\section{Conclusion and outlook}
We have constructed models in $2+1$ dimensional as well as  $3+1$ dimensional
theories in the bulk, which displays  a complete 
anisotropy in not only space and time coordinates but also among the spatial
coordinates for the $3+1$ dimensional examples. In particular, the 
exponents for the latter model comes in a specific order that is 
$z_1\geq z_2.$

If we recall from the brief review article \cite{sm}, that in one spatial 
dimension there exists an example of Tomonaga-Luttinger (TL) liquid  
which does show the quantum 
phase transition. The studies that we have done in section 2, shows that 
 probably we are moving in the proper 
direction even though we are not in a stage to directly map the gravity 
solution of  eq(\ref{config2}) to that of the TL liquid. First the TL liquid
is  a relativistic model and is studied using conformal field theory, 
whereas here we use a non-relativistic system and  with fewer symmetries 
like time translation, spatial translation and scaling symmetry. 

Second, what plays the  role of  
the dimensionless coupling $g$, from the bulk theory point of view ? 
We cannot take the Newton's constant to play
that role either in 2+1 dimension or 3+1 dimensional, as it is a dimensional 
object or $L$. 

It is certainly very interesting to  know the full criterion required to 
find either the Lifshitz point or the percolation point, so that we can have 
a better way to compute and predict things, from the bulk point of view.

The way we have approached to find the gravity dual is by demanding that the
field theory system  that we have should possess the minimal symmetry such as
time translation, spatial translation and more importantly, the scaling 
symmetry. Under the latter symmetry we want that there should be  an 
asymmetry in the way the time and the spatial coordinate transforms also 
an asymmetry between the spatial coordinates. For example, in $2+1$ dimensional
bulk theory the scaling symmetry for the temporal and spatial coordinate
\be
t\rightarrow\lambda^z~t,~~~x\rightarrow\lambda~x,~~~
\rho\rightarrow\f{\rho}{\lambda}
\ee
In $3+1$ dimensions or higher dimensions,   we can construct models for 
which there is not any asymmetry in the scaling behavior of the spatial 
coordinates like
\be
t\rightarrow\lambda^z~t,~~~x_i\rightarrow\lambda~x_i,~~~
\rho\rightarrow\f{\rho}{\lambda},
\ee
where $i$ takes the number of spatial directions. Note, in this case
we have increased the number  generators, simply due to the fact that we can 
have rotations among these spatial coordinates, $x_i$.

However, in 3+1 dimension, if we want a symmetry under scaling 
\be
t\rightarrow\lambda^{z_1}~t,~~~x\rightarrow\lambda^{z_2}~x,~~~
y\rightarrow\lambda~y,~~~\rho\rightarrow\f{\rho}{\lambda},
\ee
then we do not need to increase the number of generators beyond four 
and we  saw examples of this kind.
 
So, to summarize this paragraph, we can have a field theoretic system with 
a dual bulk theory description with the
 minimum number of symmetries like time translation, spatial translation 
and scaling with an anisotropy in  temporal and spatial coordinates (in 
2+1 dimension) as well as among all the field theory coordinates (typically in
3+1 dimension) may gives rise to the appearance of Lifshitz point.  \\

It is certainly very interesting to find the behavior of the ground state as 
we know that at the critical point,  the theory is scale 
invariant and if the ground state of these systems shows  either the 
full conformal symmetry or the Schr$\ddot{o}$dinger  group, then these
theories are being said in \cite{aff} as conformal quantum critical points 
and is suggested that a $2+1$ dimensional field theory showing all these 
at the quantum conformal critical point   must have zero resistance to 
shear stress in the two dimensional plane, which is interesting to check from 
dual bulk theory side. \\
 
{\bf Acknowledgment:} It is a pleasure to thank Shamit Kachru for  useful
and interesting correspondences and the theory group, Saha Institute of 
Nuclear Physics, Kolkata for the warm hospitality.

\end{document}